# Linking Thermal History to Shear Band Interaction and Macroscopic Ductility in Metallic Glasses

Lechuan Sun[1], Shan Zhang[1,*], Bin Xu[1], Rui Su[2], Yunjiang Wang[3,4,*], Pengfei Guan[1,5*]

[1]Ningbo Institute of Materials Technology and Engineering, Chinese Academy of Sciences, Ningbo, Zhejiang 315201, China
[2]Institute of Advanced Magnetic Materials, College of Materials & Environmental Engineering, Hangzhou Dianzi University, Hangzhou 310018, China
[3]State Key Laboratory of Nonlinear Mechanics, Institute of Mechanics, Chinese Academy of Sciences, Beijing 100190, China
[4]School of Engineering Science, University of Chinese Academy of Sciences, Beijing 101408, China
[5]Beijing Computational Science Research Center, Beijing 100193, China

**Abstract:** Shear band propagation and interaction are critical to the mechanical performance of metallic glasses and are strongly governed by thermal history, yet their microscopic mechanisms remain unclear. Here, using molecular dynamics simulations combined with a state-of-the-art annealing protocol, we systematically investigate these behaviors in a model metallic glass across effective quenching rates spanning six orders of magnitude. Through a double-notch model, we show that the normalized interaction distance—relative to the single shear-band width—is significantly larger in slowly quenched samples than in rapidly quenched ones. Atomic-scale analysis reveals that rapidly quenched samples exhibit a high density of pre-existing soft regions, which trigger correlated shear transformation zones through local vortex fields, resulting in propagation path locking and weak inter-band coupling. In contrast, slowly quenched samples exhibit enhanced structural heterogeneity and a right-shifted activation energy spectrum, promoting a single large-scale vortex field ahead of the shear band front. This field facilitates long-range stress transmission and induces shear-band deflection, convergence, and coalescence—a transition resembling a "shielding effect," in fracture mechanics, where vortex-mediated disturbances destabilize the advancing shear-band front. Our findings establish a direct microscopic connection between glass stability and shear-band-mediated plasticity and suggest that regulating shear-band interactions offers a promising route to enhance the room-temperature ductility of metallic glasses.

**Keywords**: metallic glass, shear band, STZ-Vortex model, thermal stability

# 1. Introduction

Metallic glasses have attracted considerable attention owing to their combination of high strength and hardness, as well as outstanding physical and chemical properties, including excellent wear and corrosion resistance[1-4]. However, they generally exhibit limited plasticity at room temperature, particularly under uniaxial tension, where catastrophic failure is typically triggered by the rapid propagation of one or a few dominant shear bands[5,6]. Accordingly, enhancing the room temperature ductility of metallic glasses remains a long-standing challenge in mechanics of materials.

Over the past decades, considerable efforts have been devoted to improving the plasticity of metallic glasses by controlling shear band behaviors. A widely adopted strategy is to introduce structural heterogeneities through pre-treatments such as pre-compression, surface coating, or shot peening, which generate pre-existing shear bands or other microstructural defects at, or, beneath the surface[7-9]. Upon subsequent loading, these defects can be reactivated and interact with newly formed shear bands, thereby altering the overall deformation pathway. When their orientation is aligned with the direction of maximum shear stress, pre-existing shear bands are readily reactivated, acting as localized soft regions due to the excess free volume. Conversely, when they are misaligned, they tend to impede shear band propagation, promoting the formation of multiple intersecting shear bands. Such intersections help suppress strain localization and may even induce macroscopic strain hardening. Together, these observations underscore the critical influence of shear band interactions on the plastic deformation of metallic glasses.

Extensive experimental studies have been carried out to elucidate the microscopic mechanisms of shear band interactions. Atomic force microscopy (AFM) observations by Wang et al. revealed enhanced structural stability at shear band intersections and further suggested that interacting shear bands often decompose into finer sub-bands before coalescing into new shear paths[10]. In a related study, Hu et al. examined the complex evolution of shear band morphologies—such as interweaving, deflection, and coalescence—through adjacent nanoindentation experiments[11]. Nevertheless, the spatial and temporal resolution of current experimental techniques remains insufficient for directly capturing these processes at atomic level. Consequently, molecular dynamics (MD) simulations have become a valuable complementary approach, providing atomistic insights into the evolution of shear band.

In recent work, Sopu et al.[12] proposed a inspiring shear transformation zone-vortex (STZ-Vortex) self-catalysis model, derived from large-scale MD simulations, to describe the atomic scale structural units governing the nucleation, propagation, and interaction of shear bands. They identified an alternating sequence of shear STZs and vortex-like rotational zones within the shear deformation region, which can self-organize along specific directions to form stable continuous shear paths. As a result, the mechanism provides a coherent explanation for shear localization, shear band branching, and multi-band interactions, offering new insights into the atomistic dynamics of shear band evolution. Furthermore, the model shows promise for elucidating structural relaxation processes in metallic glasses. It is important to note that most conventional MD simulations employ quenching rates that are at least five orders of magnitude faster than those achievable in laboratory fabrication[13,14]. Consequently, MD-generated metallic glasses often display increased apparent ductility, arising from the extra free volume due to rapid quenching. Extra free volume facilitates nucleation and propagation of shear bands, potentially biasing the observation of shear banding behaviors.

Recent studies have shown that the swap Monte Carlo (Swap MC) algorithm can produce ultrastable multi-component glasses with energy state comparable to those of experimental metallic glasses[15,16]. Building on these advances, our previous work developed an efficient annealing strategy—the hybrid thermal cycling (HTC) method—which integrates MD and Monte Carlo techniques to achieve an ultralow effective cooling rate of approximately $10^4$ K/s[17]. This unprecedented slow quenching rate is comparable to that obtained in typical copper mold casting experiments, thereby narrowing the gap between atomistic modeling and experiment. Using this HTC strategy, our previous work clarified the size-dependent fracture behavior of metallic glass nanowires, showing that thermal history and sample dimensions jointly dictate the transition of fracture mode from shear banding to necking[18]. In parallel, Ding et al. employed a similar hybrid MD/MC approach to investigate the deformation of ultrastable metallic glasses. They clarified that thermal stability fundamentally alters the shear band propagation mechanism: hyper-quenched glasses exhibit intermittent, STZ-mediated "stop-and-go" propagation interspersed with vortex-like rotation, whereas slowly cooled stable glasses undergo more continuous shear softening accompanied by larger dimension of vortex fields. These results firmly established how shear band nucleation and propagation are dependent on structural state of glasses[19,20]. Despite these advances, most studies have concentrated on single shear band. It is rare to see report on the collective dynamics of multiple interacting bands in ultrastable glasses, including the band deflection, branching, coalescence, as well as path competition. Since collective dynamics arising from interaction between shear bands are critical for understanding and ultimately controlling macroscopic plasticity, here we systematically investigated multiple shear bands propagation and interaction in glass samples with distinct thermal history. We are particularly interested in exploring the true STZ-Vortex synergetic mechanism in an experimentally relevant metallic glass sample.

## 2. Method

We performed MD simulations using the Large-scale Atomic/Molecular Massively Parallel Simulator (LAMMPS) package[21] to investigate the propagation and interaction of shear bands and their underlying microscopic mechanisms. Specifically, we focused on a well-studied model metallic glass, $Zr_{50}Cu_{40}Al_{10}$, with interatomic interactions described by the embedded-atom method (EAM) potential developed by Cheng et al., which has been shown to accurately reproduce the structural, thermodynamic, and mechanical properties of this amorphous alloy[22].

A small slow quenching rate sample was first prepared using the HTC method, yielding an effective cooling rate of approximately $10^4$ K/s judged from the empirical extrapolation of the MD cooling history[17]. The initial configuration was a cubic cell of 5 nm × 5 nm × 5 nm containing approximately 8000 atoms, with periodic boundary conditions applied in all three directions. During the HTC process, the swap Monte Carlo (Swap MC) algorithm was employed, performing 10 atomic exchange attempts at each MD step, while the system temperature was controlled via velocity rescaling[17]. For comparison, a rapid quenching rate sample was prepared using a conventional rapid cooling procedure: the same sized configuration was first fully melted at 2000 K and 0 GPa, then quenched to 100 K at a rate of $10^{10}$ K/s, followed by a 2 ns relaxation to ensure thermal stability. The small cubic cell was replicated eight times along the y-direction and sixteen times along the z-direction, resulting in a simulation box with dimensions of 5 nm × 40 nm × 80 nm and containing approximately 1024000 atoms. Two identical notches were introduced at different positions on opposite sides of the sample. The axial distance $D$ between the notches was varied from 0 to 40 nm to control the spacing $S$ between the potential shear bands and thereby tune their interaction strength (see Figure 1a for geometry). In practice, the actual spacing $S$ does not exactly equal the geometric projection $D / 2$, because the shear bands may deviate from the ideal 45° orientation (maximum shear stress direction) with respect to the loading axis, and their nucleation points may not coincide precisely with the notch edges[23].

To systematically examine the shear band response under different structural states, we performed uniaxial tensile simulations on all samples. Before loading, each configuration was relaxed at 100 K for 1 ns to release local residual stresses. A uniaxial tensile strain was then applied along the y-direction at a constant strain rate of $4\times10^7$ /s. The boundary conditions were set as free along the y-direction and periodic along the x- and z-directions to accommodate lateral deformation. All simulations were conducted in the NPT ensemble with a time step of 2 fs. To ensure statistical reliability, multiple independent double-notch samples (generated with different initial velocity seeds) were prepared for selected key values of $D$, while single samples were used for other values to verify general trends. In total, approximately 70 rapid quenching rate and 80 slow quenching rate samples were produced. For clarity, rapid and slow quenching rate samples are denoted as "H-Sample" and "L-Sample" throughout the figures and discussion.

## 3. Results

Figure 1b shows representative stress-strain curves for the L-Sample and H-Sample, respectively, under a shear band spacings $S \approx 20$ nm. All samples exhibit the characteristic tensile response of metallic glasses: an initial linear elastic regime, a nonlinear transition near the yield point, a post-yield stress drop, and a subsequent steady-state flow stage. The L-Sample exhibits larger elastic modulus, higher yield strength ($\tau_y$) and relatively larger critical strain of yield, reflecting its denser atomic packing and higher degree of short-range order (SRO). These structural features suppress the activation of STZs, thereby increasing resistance to plastic deformation. By contrast, the H-Sample contains larger content of excess free volume, which promotes shear initiation and results in a lower overall strength[24-26]. After reaching the yield point $\tau_y$, both the H-Sample and L-Sample exhibit a pronounced stress drop before stabilizing at a nearly constant stress level $\tau_s$, commonly referred to as the flow stress[27,28]. Here, we define a normalized stress-drop parameter $\Delta\tau / \tau_y$, where stress overshoot $\Delta\tau = \tau_y - \tau_s$, to quantify the extent of post-yield instability, which has been shown to be equivalent to local strain correlation in indicting level of strain localization[24]. A smaller $\Delta\tau / \tau_y$ corresponds to a more gradual stress relaxation following yield and thus a greater capacity to accommodate plastic strain[29,30].

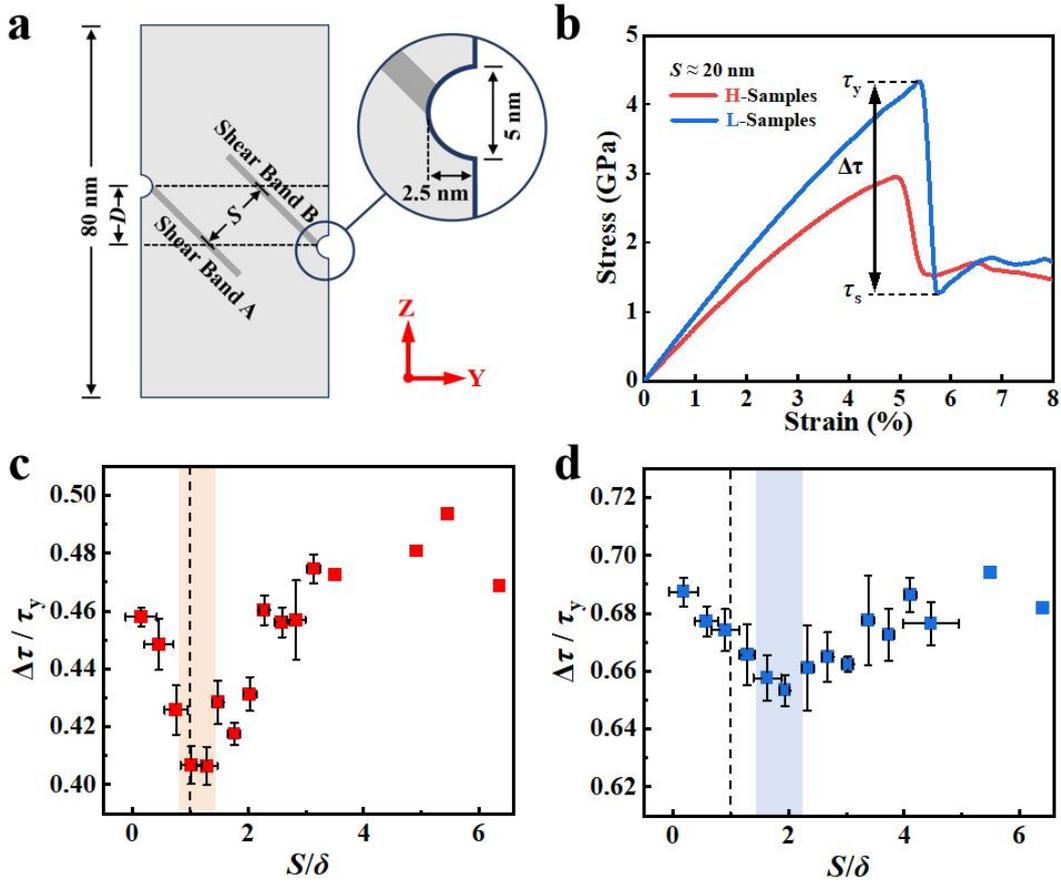

**Fig. 1.** Effect of notch spacing $S$ on the shear-banding-driven mechanical response of different glass samples. (a) Geometry of the double-notch sample and locations of two shear bands. The inset enlarges the notch region; gray bands mark shear bands formation upon tension. (b) Stress-strain curves of the H-Sample and L-Sample with shear band spacing $S \approx 20$ nm. (c, d) Variation of the normalized stress drop $\Delta\tau / \tau_y$ as a function of the scaled spacing $S/\delta$, for H-Sample and L-Sample, respectively. $\delta$ denotes the characteristic width of a fully developed shear band. Error bars represent the standard error.

Figures. 1c and 1d present the variation of $\Delta\tau / \tau_y$ as a function of the scaled spacing $S/\delta$, for H-Sample and L-Sample, where $\delta$ denotes the characteristic width of a fully developed shear band in each sample (see Supplementary Material for details)..

For critical spacing intervals, multiple simulations were conducted using identical initial configurations to ensure statistical reliability, while single simulations were performed for non-critical cases. The shaded regions indicate the $S/\delta$ ranges exhibiting reduced $\Delta\tau / \tau_y$, corresponding to improved plastic response. A clear non-monotonic trend is observed: within a certain range of $S/\delta$, increasing the spacing reduces stress overshoot and therefore enhances plastic plasticity, whereas further increase in spacing causes a gradual enhancement on stress over and encourages brittle failure. This behavior directly reflects the influence of inter-band interaction strength on the plastic response. The vertical dashed lines mark the typical widths of fully developed shear bands in the two samples (see Supplementary Material for details). Since the mature bands in the rapidly quenched glass are generally wider than those in the slowly cooled glass[17,31], one might naively expect the effective interaction range, when expressed in terms of the absolute spacing $S$, to be larger in the H-Sample. Under this simple geometric expectation, the spacing associated with maximal plasticity enhancement should therefore appear at larger $S$ in H-Sample. However, the opposite trend is observed: the minimum in $\Delta\tau / \tau_y$ occurs at noticeably larger $S$ in the slowly cooled sample. This counterintuitive spacing dependence suggests that the effective interaction length is governed by factors beyond simple geometric considerations, a point examined further in the following sections.

To further illustrate this effect of notch spacing on shear banding, Figure 2 displays the shear band morphologies near the yielding point for representative samples with $S \approx 0$ nm, 3 nm, 8 nm, and 20 nm. Atomic shear strain was calculated for each atom by averaging over its neighbors within a cut-off radius of 10 Å[32]. The same averaging scheme is applied to all related parameters discussed hereafter.

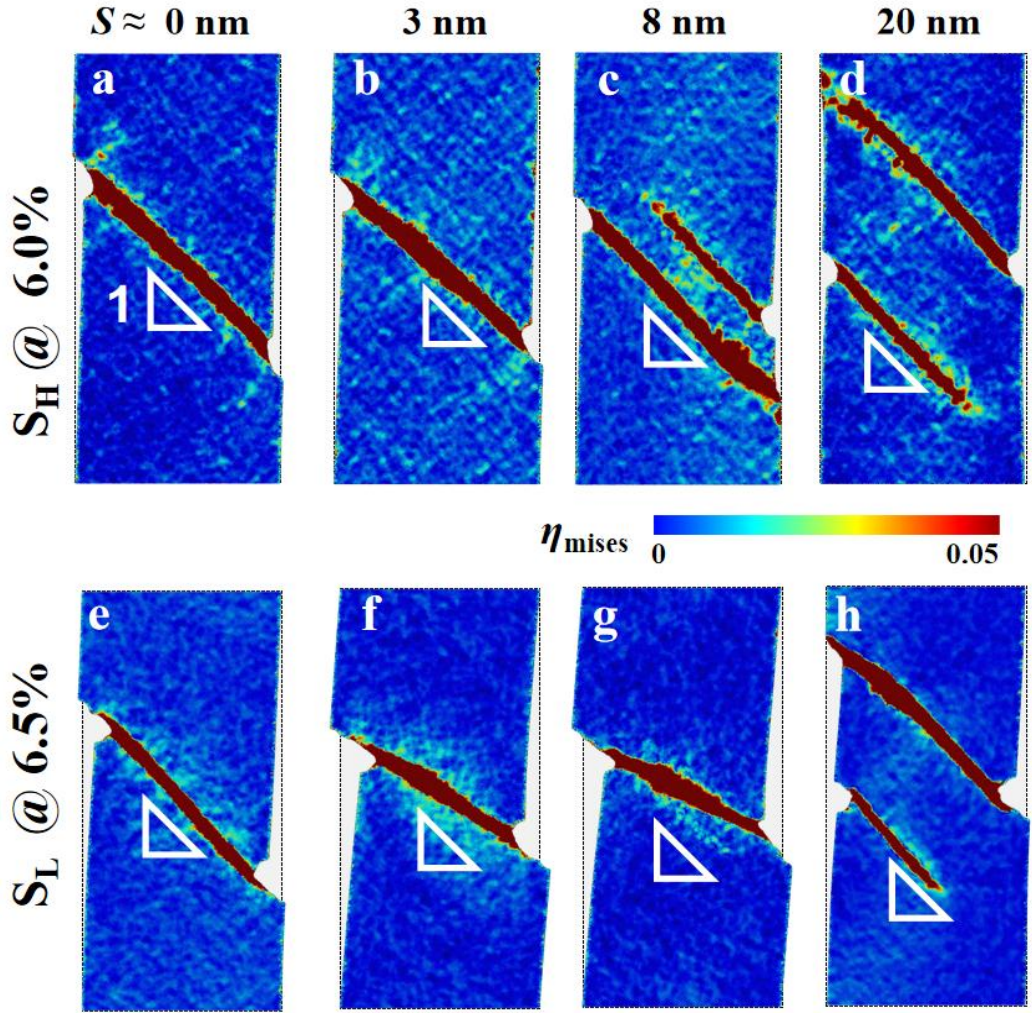

**Fig. 2.** Influence of shear band spacing $S$ on shear banding interaction in the H-Sample (a-d) and L-Sample (e-h). All snapshots are taken at an engineering strain of approximately $\varepsilon \approx 6\%$ for the H-Sample and $\varepsilon \approx 6.5\%$ for the L-Sample (slightly beyond the yield point). The color scale indicates the magnitude of local atomic shear strain, estimated from consecutive images with a time interval of 100 ps.

When the spacing $S$ approaches zero, the two shear bands on the opposite sites of sample essentially follow the same trajectory, resulting in a deformation pattern nearly identical to that of a single shear band. Since no additional energy dissipation mechanisms are introduced, the plastic performance shows no noticeable improvement (Figures 2a and 2e), based on the level of stress overshoot. Moreover, the intensified stress concentration under this condition facilitates shear instability, causing slight reductions in both yield strength and yield strain. As the spacing increases, coupling effects between the shear bands gradually emerge. At small but finite spacings, strong mechanical coupling develops at the shear band fronts, leading to partial coalescence and the formation of a broadened shear zone, maintaining the morphology of single shear band. This interaction enhances energy dissipation and promotes strain delocalization (band thickening), thereby improving overall plasticity. At larger spacings, however, the interaction between shear bands becomes weak, and the two paths propagate independently without merging or reconstructing their trajectories. The reduced interaction recovers the scenario of single shear band and lowers strain delocalization, which results in a slight increase in stress overshoot and decrease in plastic deformability compared with the small spacing case.

To gain further insight into these trends, we next examine the detailed deformation morphologies with double shear bands from notches at representative spacings. Figures 3 and 4 show the temporal evolution of shear strain in an enlarged view of the shear band interaction region for the H-Sample and L-Sample at predefined spacings of $S \approx 3$ nm and $S \approx 8$ nm. Dashed lines indicate the shear band propagation direction, and white circles mark STZs activated during shear band propagation. There are necessary rotation regions between STZs to accommodate shear band propagation. The comparison highlights the distinct mechanisms of shear band interaction and coalescence in different spacing that arise from the two distinct structural states.

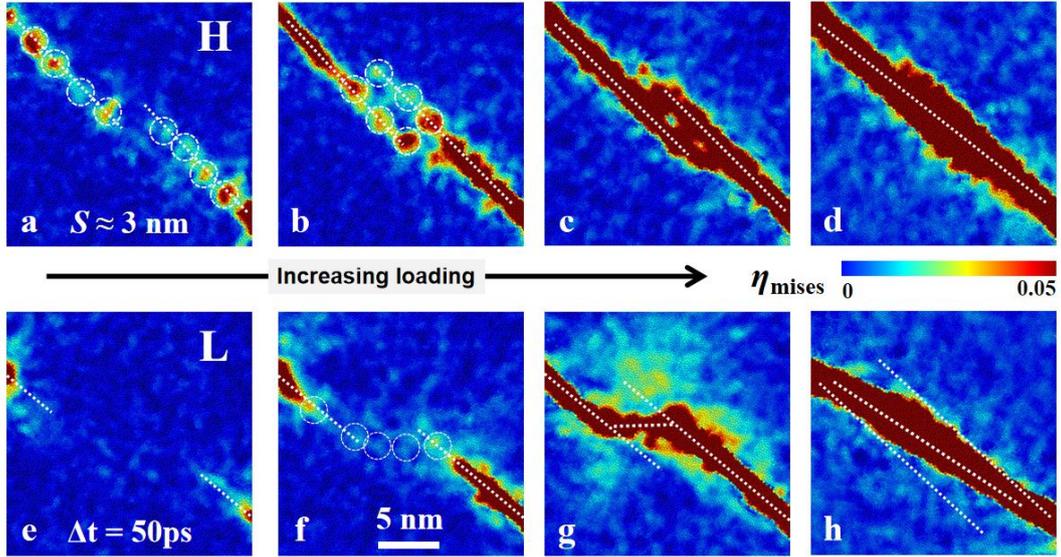

**Fig. 3.** Evolution of atomic von Mises strain $\eta_{mises}$ during tensile loading for H-Sample (a-d) and L-Sample (e-h) at a fixed shear band spacing of $S \approx 3$ nm.

For the H-Sample, the strain patterns reveal that there are continuously activated STZs (approximately 4-5) in the band front, which tend to nucleate preferentially in soft regions separated by rotations. This mechanism results in a highly discontinuous pattern composed of several isolated zones. These STZs align sequentially along the shear direction and gradually connect to form a stable strain channel, a phenomenon consistent with previous MD simulations[33]. It promotes early-stage coalescence of the two shear bands in $S \approx 3$ nm sample and is accompanied by significant local strain energy release and pronounced shear band broadening. The two mechanisms enhance plastic deformation and improve mechanical stability upon subsequent loading. Consistently, the stress-strain response in Figure S1 shows that the H-Sample with $S \approx 3$ nm exhibits a more gradual post-yield stress drop compared with the steeper softening observed at spacings such as $S \approx 0$ nm and $S \approx 20$ nm.

Under larger spacing $S$ between shear bands, the interaction gradually weakens. As shown in Figure 4, when $S \approx 8$ nm, the two shear bands in the H-Sample evolve along separated and well-defined paths, each maintaining a stable propagation direction without deflection. Although some STZs are activated in the intersection region deviating from the major band direction, larger STZs are preferentially triggered along the shear band direction, which promotes continued propagation along this path. Eventually, a dominant band propagation path emerges and continues to develop, while the other premature shear band ceases to grow due to stress release. Throughout the process, the shear bands exhibit a pronounced path locking behavior, resulting in two fully separated shear bands without any merging event. This indicates that the shear deformation becomes directionally guided along a dominant path at an early stage and subsequently maintains its directional stability by continuously activating STZs along the predetermined propagation direction, making it difficult for the system to undergo path switching or merging. Therefore, when the notch spacing is large enough, multiple shear bands evolves independently, and the mechanical response resembles the case of a single shear band.

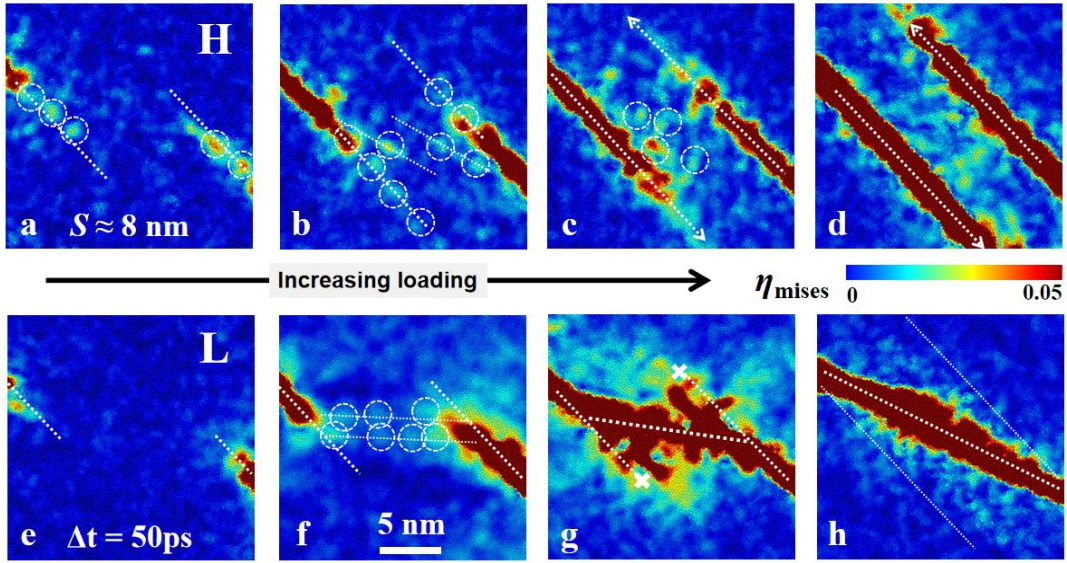

**Fig. 4.** Evolution of atomic von Mises strain $\eta_{mises}$ during tensile loading for H-Sample (a-d) and L-Sample (e-h) at a fixed shear band spacing of $S \approx 8$ nm.

In contrast, the L-Sample lacks pre-existing soft regions at the shear band front, and therefore the STZ nucleation events are strongly suppressed (often only a single STZ is triggered, or none at all), as shown in Figures 3e and 4e. In other words, the deformation is well confined in a local region by the hard structure in the L-Sample. Consequently, shear band advances mainly through a shear-softening mechanism, making its propagation path highly sensitive to local microstructural perturbations. In structurally homogeneous regions, the shear band generally follows the direction of maximum shear stress, but it becomes more prone to deflection when encountering perturbations in the local stress field or variations in atomic packing density. As the two shear bands gradually approach, several small STZs are intermittently activated along the line connecting their tips and progressively link into a discontinuous yet directional strain channel. This incipient strain channel guides the front of one shear band toward the other, leading to eventual coalescence, as illustrated in Figures 3g and 4g. Compared to the unstable glasses, multiple neighboring shear bands pretend to merge in ultrastable glass due to stronger structural confinement to plastic strain.

According to the STZ-Vortex model[12], activation of a STZ generates an Eshelby-like quadrupolar elastic field, characterized by compressive and tensile lobes oriented along orthogonal directions. This antisymmetric field drives collective, vortex-like rotations of the surrounding atoms, clockwise along the shear front and counterclockwise in the perpendicular direction. Shear band formation can thus be described as an autocatalytic sequence of two coupled structural units: STZs and localized vortex-like rotational zones. Their repeated alternation promotes long range STZ percolation and establishes the dominant propagation path[12,33-36]. This framework provides a useful basis for interpreting the atomic scale rotational displacement patterns observed in our simulations and facilitating rationale of shear band interaction.

Figures 5 and 6 show the temporal evolution of the rotation angle $\theta$ in an enlarged view of interaction region for the H-Sample and L-Sample. The color map indicates the direction and magnitude of atomic rotation (red for clockwise rotation, and blue for counterclockwise rotation). Dashed lines denote the shear band propagation direction, while black and white circles mark the locations of local rotation cores emerging during propagation. For the H-Sample with $S \approx 3$ nm, the rotation angle maps reveal a periodic array of vortex-like structures at the shear band front, which self-organize into a coherent rotational bridging zone along the propagation path, as shown in Figures 5a and 6a. This reflects a highly coordinated rotational response, where vortices organize into a persistent rotational bridging zone that supports stable, band-like propagation of the shear front. At larger spacing ($S \approx 8$ nm), the interaction weakens both in spatial extent and in rotational coherence, as shown in Figures 6(a-d). Although localized vortical structures are still present, they remain disconnected and fail to form a bridging zone. With continued loading, several incipient pre-shear trajectories emerge as dispersed, localized vortical disturbances between the shear bands, yet they never consolidate. Instead, the shear bands propagate along distinct, well-defined paths and remain fully separated, with no coalescence or pathway reconstruction throughout deformation.

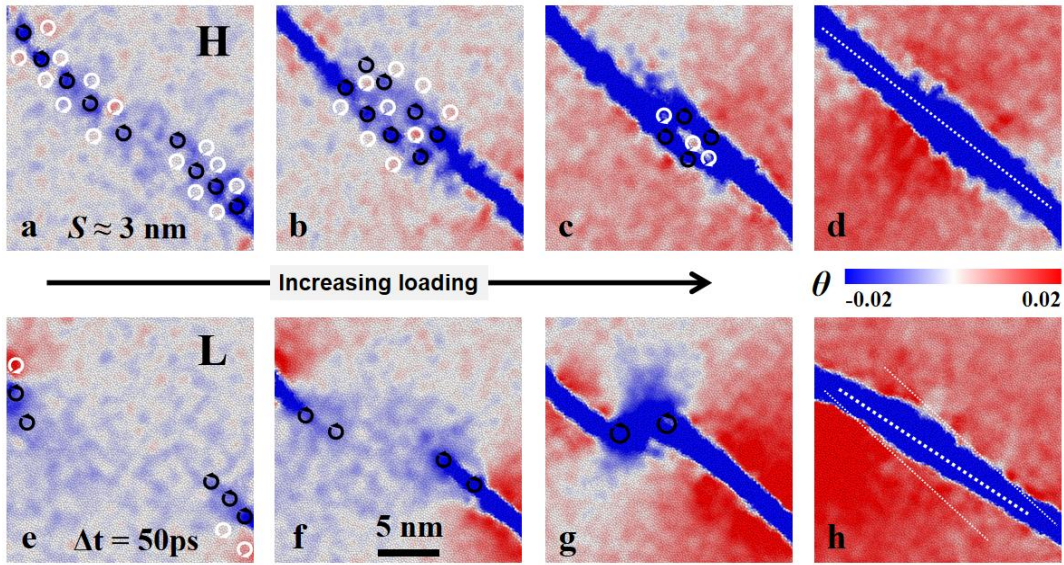

**Fig. 5.** Evolution of atomic rotation angle $\theta$ during tensile loading for H-Sample (a-d) and L-Sample (e-h) at a fixed shear band spacing of $S \approx 3$ nm.

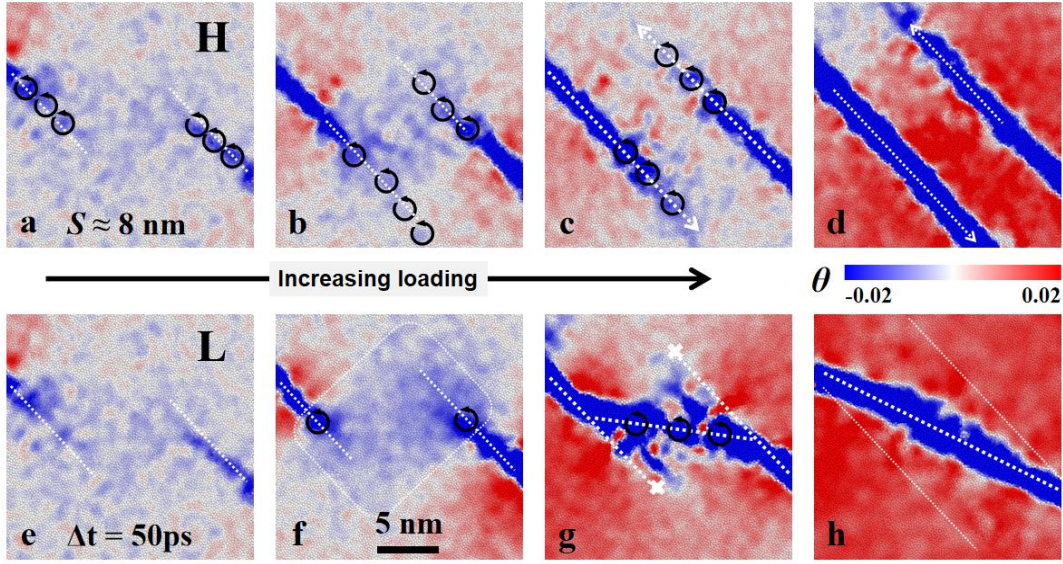

**Fig. 6.** Evolution of atomic rotation angle $\theta$ during tensile loading for H-Sample (a-d) and L-Sample (e-h) at a fixed shear band spacing of $S \approx 8$ nm.

In contrast, for the L-Sample, the shear band front exhibits a coherent, large-scale vortex structure in the region between the two shear bands, rather than multiple discrete vortices. The corresponding rotation angle distribution exhibits a well-defined vortex-like pattern, with minimal fluctuations in both rotation angle and magnitude, as shown in Figures 5e and 6e. This suggests a more uniform microscopic deformation mechanism ahead of the shear front, characterized by collective atomic motion over extended spatial scales compared with the H-Sample. Such an organized rotational response can exert a long-range influence on the surrounding stress and deformation fields, effectively steering the shear front toward the neighboring shear band. As loading progresses, the front of one shear band undergoes a pronounced deflection toward the other and eventually merges with it, even at a spacing as large as $S \approx 8$ nm (Figures 6e-h).

## 4. Discussion

In the double-notch configuration, the presence of two interacting shear bands generates a complex stress field, making it difficult to isolate the specific influence of the rotation field on their propagation paths. To eliminate such interference, we constructed a single-notch model under a more controlled loading configuration. This configuration allows us to examine the causal relationship between the rotation field distribution and shear band trajectory selection and propagation stability, and to assess the applicability and generality of the aforementioned mechanism. Figures 7b and c illustrate the von Mises strain associated with the nucleation and early growth of the shear band under tensile loading. In the H-Sample, early-stage deformation involves the activation of several discrete STZ clusters ahead of the notch, roughly aligned along two symmetric directions. These clusters subsequently coalesce into a continuous shear band, accompanied by a coherent vortex-like rotational field. The shear band propagates at an angle close to the maximum shear stress direction (approximately 45°), reflecting the alignment of its trajectory with the dominant loading induced stress orientation. In the L-Sample, strain rapidly concentrates near the notch tip and develops into a shear band that propagates across the sample via a shear-softening mechanism. Rotation activity is largely confined to the notch vicinity, where it organizes into a larger scale vortex structure rather than forming a spatially continuous band. The shear band propagates at an angle of about 41°, slightly different from that in the H-Sample. This difference can be interpreted within the STZ-Vortex model[12]: the larger ratio of tensile to compressive forces acting on the STZs biases their activation toward the tensile stress component, resulting in a modest deviation in the observed angle compared with the H-Sample[37,38]. Figure S4 show the atomic shear strain and rotation angle distributions at the shear band front, together with their normalized profiles along the propagation path. In the H-Sample, both quantities exhibit a distinct decaying peak-valley pattern, indicating a spatially continuous deformation channel. In contrast, the L-Sample profiles are smoother with smaller fluctuations, providing no clear directional signature for the advancing front.

Overall, these results demonstrate that the structural state of the glass strongly influences both the initiation and evolution of shear band interactions. In the hyper-quenched sample, abundant pre-existing soft regions promote frequent STZ activations and the development of continuous strain and rotational bridging zones, leading to early coalescence at small spacings and path locking at larger spacings. In the slowly cooled sample, STZ activity is significantly suppressed, but the shear band front is more sensitive to local microstructural variations, enabling deflection and eventual coalescence even at relatively large spacing. As shown in Figure S3, the rotation field analysis further reveals that the difference originates from fundamentally different manifestations of the underlying quadrupolar field, appearing as periodic, spatially correlated vortices in the hyper-quenched glass and as large-scale, coherent rotations in the slowly cooled glass, which ultimately governs the forms of shear band interactions.

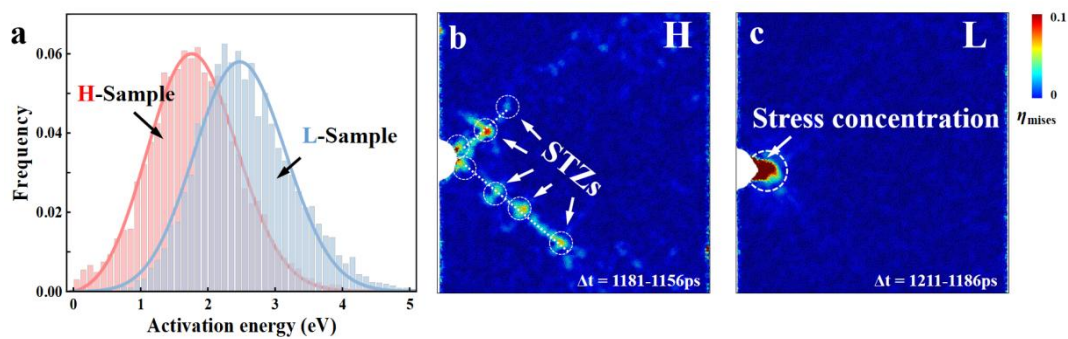

**Fig. 7.** (a) Distribution of activation energies for the H-Sample (red) and L-Sample (blue), with Rayleigh-like fits shown as solid curves. (b, c) Atomic von Mises strain fields during shear-band nucleation and early growth near the notch in the H-Sample and the L-Sample, respectively.

The difference in shear band interaction can be further understood by examining the activations on the potential energy landscape at atomic scale[39-41]. As shown in Figure 7a, we employed the activation-relaxation technique to quantify the distribution of activation energies of the H-Sample and L-Sample[42,43]. The L-Sample exhibits a broader and upward-shifted distribution, with a peak near 2.5 eV, indicative of a more structurally stable local environment. In such a landscape, shear band propagation is more likely to encounter large energy barriers, giving rise to more heterogeneous propagation modes, including deflections, interruptions, and directional changes[44]. In contrast, the H-Sample exhibits a narrower distribution skewed toward lower energies, with a peak around 1.7 eV. Therefore, the dynamic heterogeneity is reduced, and the fertile regions are more abundant. This low-barrier landscape facilitates the correlated activation of multiple STZs, allowing strain to percolate at reduced energetic cost and enabling smoother advancement along the initial path[45,46].

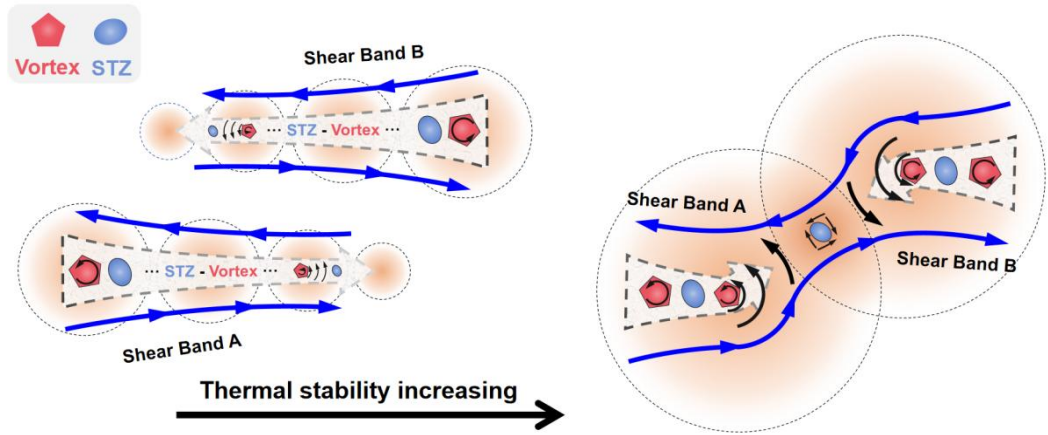

**Fig. 8.** Schematic illustration of the shear band interaction mechanisms under different thermal stabilities. Pentagons denote vortex-like rotational zones, Ellipses represent STZs, and blue arrows indicate the propagation direction. Shaded regions indicate the extent of the stress field.

The present results demonstrate that the thermal stability of the glass, governed by its thermal history, fundamentally alters the interaction dynamics between neighboring shear bands in metallic glasses. At low thermal stability, the glass retains substantial nanoscale structural heterogeneity, producing an intermixed distribution of "soft" and "hard" regions[47,48]. This structural landscape promotes spatially selective and fragmented STZ activation. At the shear band front, STZs are triggered in a correlated manner, progressively linking pre-existing soft regions into a continuous deformation channel. Even when multiple shear bands evolve concurrently, their interaction is constrained by the underlying activation mechanism. This is attributable to two factors: (i) correlated STZ activation locks the shear trajectory along predefined soft channels, thereby suppressing directional deviation; and (ii) the stress field remains spatially confined because most of the local stress is consumed by STZ activation, leaving little available for long range transmission. In contrast, at high thermal stability, the glass exhibits a higher degree of structural homogeneity, characterized by the absence of pronounced soft regions, a generally higher elastic modulus, and a greater prevalence of five-fold symmetry. These features produce a propagation mechanism dominated by shear softening at the shear band front, accompanied by large-scale cooperative atomic displacements in both shear and rotation[47]. In this mode, the advancing front generates an extensive vortex-like displacement field together with broad stress concentration zones. Unlike at low thermal stability, the stress concentration is not locally dissipated via discrete STZ formation but instead persists and expands. This enables effective long range stress field percolation between neighboring shear bands, whereby the stress or displacement field from one advancing shear band can remotely perturb another incipient or evolving shear band, triggering its premature activation or deflection toward the high stress region, as illustrated in Figure 8. In practical terms, such long-range coupling often manifests as shear band deflection, convergence, and eventual coalescence into a unified primary shear plane. The merging of initially separate shear bands provides direct evidence for vortex-mediated coupling, leading to the formation of a single dominant shear band. Even at larger spacings, induced deflection remains evident, underscoring the far field impact of vortex-mediated stress transmission on shear band trajectories. These distinct propagation regimes ultimately shape the form of shear band interactions and, consequently, influence plastic deformability of metallic glasses.

The mechanistic distinction outlined above signals a transition from a glass-specific description to a broader mechanical framework governing front stability under perturbing fields. Classical atomistic studies of crack propagation in crystalline metals have long established that the stress field near a crack tip is highly sensitive to additional perturbing fields[49,50]. In these systems, the emission of dislocations introduces secondary stress components that modify the local mode-I and mode-II fields, thereby altering the stability of the tip stress concentration[51,52]. This class of behavior is widely referred to as a shielding effect, whose defining characteristic is not merely a reduction of the local driving stress, but a disruption of the directional singularity that normally stabilizes the advance of the front. A closely related mechanical phenomenon emerges in the L-Sample metallic glass examined here, even though the underlying defect physics differs fundamentally from dislocation activity in crystalline metals. Nonetheless, the macroscopic mechanism through which perturbing fields destabilize the advancing front remains analogous. The shear band front in the L-Sample constitutes a sharp, localized region of intense shear softening, analogous in function to a crack tip in determining the directionality of deformation advance. In addition, unlike the highly localized, STZ-dominated front in the hyper-quenched H-Sample, the front in the L-Sample is embedded within a large-scale, spatially coherent vortex field. This vortex field generates long-range rotational displacements that act as an extended perturbing stress field, directly analogous to the secondary stress fields produced by dislocations in classical shielding. It reshapes the local stress landscape that would otherwise maintain a well-defined propagation direction. Therefore, the front in L-Sample becomes susceptible to systematic deviations driven by the spatial gradients imposed by the vortex field. It may rotate away from its initial orientation, shift toward regions of compatible stress alignment, or even be drawn toward the front of a neighboring shear band, representing the exact mechanical signatures expected when perturbing stress fields act to reshape a crack-tip singularity in the classical shielding literature[53,54]. Thus, although the microscopic origins differ (collective vortex rearrangements versus dislocation emission), the mechanical outcome is analogous: the propagation front is no longer governed by a single dominant stress direction, but by a continuously evolving stress landscape shaped by nearby perturbations. This contrast maps cleanly onto the classical shielding framework: the H-Sample behaves as a non-shielded system, where STZ-mediated plasticity dissipates stress locally and preserves a sharply defined propagation direction. The L-Sample, by contrast, exhibits the mechanical signature of shielding, as the vortex-generated perturbing stress fields destabilize the front and promote deflection, reorientation, and interaction with neighboring bands.

## 5. Conclusion

In conclusion, this study clarifies how quenching rate controls the atomic-scale structural state of metallic glasses and thereby governs the dominant mechanisms of shear band propagation and interaction. Molecular dynamics simulations of $Zr_{50}Cu_{40}Al_{10}$, prepared via conventional hyper-quenching and experimentally comparable slow-quenching protocols, reveal two distinct regimes of shear band interaction. In the hyper-quenched glass, abundant pre-existing soft regions promote spatially correlated activation of shear transformation zones along predefined channels, resulting in path-locked shear bands that resist deflection or merging. Conversely, in the slowly cooled glass, structurally homogeneous shear band fronts generate large-scale cooperative vortex fields, enabling long-range stress transmission, shear band deflection, and eventual coalescence even at relatively wide spacings. This loss of directional stability closely resembles the shielding-like front perturbations observed in fracture mechanics. These findings establish shear band interaction as a tunable microstructural process, offering a direct route to tailor deformation pathways and stabilize plastic flow in metallic glasses. The link between atomic-scale mechanisms and macroscopic response provides a foundation for processing strategies that exploit interaction-mediated plasticity in these nominally brittle amorphous materials.

## Acknowledgements

Lechuan Sun thanks Heng Kang and Yicheng Wu for their helpful discussions. This work was supported by the National Natural Science Foundation of China (Grants No. T2325004 and 52161160330), the Advanced Materials-National Science and Technology Major Project (Grant No. 2024ZD0606900), and the Talent Hub for "AI+ New Materials" Basic Research. Yunjiang Wang is supported by the Strategic

Priority Research Program of Chinese Academy of Sciences (Grants No. XDB0510301 and No. XDB0620103) and the National Natural Science Foundation of China (Grant No. 12472112). We acknowledge the computational support from the Beijing Computational Science Research Center (CSRC).

# References

1    Ashby, M. F. & Greer, A. L. Metallic glasses as structural materials. *Scr. Mater.* **54**, 321-326 (2006).

2    Johnson, W. L. Bulk glass-forming metallic alloys: Science and technology. *MRS Bull.* **24**, 42-56 (1999).

3    Inoue, A., Shen, B., Koshiba, H., Kato, H. & Yavari, A. R. Cobalt-based bulk glassy alloy with ultrahigh strength and soft magnetic properties. *Nat. Mater.* **2**, 661-663 (2003).

4    Wang, W. H., Dong, C. & Shek, C. Bulk metallic glasses. *Mater. Sci. Eng. R Rep.* **44**, 45-89 (2004).

5    Sha, Z., Teng, Y., Poh, L. H., Wang, T. & Gao, H. Shear band control for improved strength-ductility synergy in metallic glasses. *Applied Mechanics Reviews* **74**, 050801 (2022).

6    Luo, J., Huang, L., Shi, Y. & Deng, B. The dynamics of shear band propagation in metallic glasses. *Acta Mater.* **248**, 118787 (2023).

7    Zhou, C., Zhang, H., Yuan, X., Song, K. & Liu, D. Applicability of pre-plastic deformation method for improving mechanical properties of bulk metallic glasses. *Materials* **15**, 7574 (2022).

8    Cao, Q.-P. *et al.* Effect of pre-existing shear bands on the tensile mechanical properties of a bulk metallic glass. *Acta Mater.* **58**, 1276-1292 (2010).

9    Nieh, T., Yang, Y., Lu, J. & Liu, C. T. Effect of surface modifications on shear banding and plasticity in metallic glasses: An overview. *Progress in Natural Science: Materials International* **22**, 355-363 (2012).

10   Lu, Y. *et al.* Shear-banding induced indentation size effect in metallic glasses. *Scientific Reports* **6**, 28523 (2016).

11   Huang, H. & Yan, J. Investigating shear band interaction in metallic glasses by adjacent nanoindentation. *Mater. Sci. Eng. A* **704**, 375-385 (2017).

12   Şopu, D. STZ-Vortex model: the key to understand STZ percolation and shear banding in metallic glasses. *J. Alloys Compd.* **960**, 170585 (2023).

13   Wang, W. H., Lewandowski, J. & Greer, A. L. Understanding the glass-forming ability of Cu50Zr50 alloys in terms of a metastable eutectic. *J. Mater. Res.* **20**, 2307-2313 (2005).

14   Yang, G. N., Shao, Y., Yao, K. F. & Chen, S. Q. A study of cooling process in bulk metallic glasses fabrication. *AIP Advances* **5**, 117111 (2015).

15   Ninarello, A., Berthier, L. & Coslovich, D. Models and algorithms for the next generation of glass transition studies. *Phys. Rev. X* **7**, 021039 (2017).

16   Ozawa, M., Berthier, L., Biroli, G., Rosso, A. & Tarjus, G. Random critical point separates brittle and ductile yielding transitions in amorphous materials. *Proc. Natl. Acad. Sci. USA* **115**, 6656-6661 (2018).

17   Su, R. *et al.* Atomic origin of annealing embrittlement in metallic glasses. *arXiv:2208.13747 [cond-mat.mtrl-sci] (2022)*

18   Sun, L., Zhang, S., Su, R., Wang, Y. & Guan, P. Size-and stability-dependent fracture scaling in nanoscale metallic glass. *Acta Mater.* **292**, 121046 (2025).

19   Yu, J. *et al.* Structural state governs the mechanism of shear-band propagation in metallic glasses. *Proc. Natl. Acad. Sci. USA* **122**, e2427082122 (2025).

20   Zhang, Z., Ding, J. & Ma, E. Shear transformations in metallic glasses without excessive and predefinable defects. *Proc. Natl. Acad. Sci. USA* **119**, e2213941119 (2022).

21   Plimpton, S. Fast parallel algorithms for short-range molecular dynamics. *J. Comput. Phys.* **117**, 1-19 (1995).

22   Cheng, Y. Q., Ma, E. & Sheng, H. W. Atomic level structure in multicomponent bulk metallic glass. *Phys. Rev. Lett.* **102**, 245501 (2009).

23   Lund, A. & Schuh, C. The Mohr–Coulomb criterion from unit shear processes in metallic glass. *Intermetallics* **12**, 1159-1165 (2004).

24   Wei, D., Wang, Y.-J. & Ogata, S. Statistical structural mechanism for shear localization in amorphous materials. *Acta Mater.* **298**, 121385 (2025).

25   Lee, M., Lee, C.-M., Lee, K.-R., Ma, E. & Lee, J.-C. Networked interpenetrating connections of icosahedra: Effects on shear transformations in metallic glass. *Acta Mater.* **59**, 159-170 (2011).

26   Cheng, Y., Cao, A. J., Sheng, H. & Ma, E. Local order influences initiation of plastic flow in metallic glass: Effects of alloy composition and sample cooling history. *Acta Mater.* **56**, 5263-5275 (2008).

27   Feng, S. *et al.* Atomic structure of shear bands in $Cu_{64}Zr_{36}$ metallic glasses studied by molecular dynamics simulations. *Acta Mater.* **95**, 236-243 (2015).

28   Feng, S. *et al.* A molecular dynamics analysis of internal friction effects on the plasticity of $Zr_{65}Cu_{35}$ metallic glass. *Materials & Design* **80**, 36-40 (2015).

29   Peng, C. *et al.* Deformation behavior of designed dual-phase CuZr metallic glasses. *Materials & Design* **168**, 107662 (2019).

30   Kim, H.-K., Lee, M., Lee, K.-R. & Lee, J.-C. How can a minor element added to a binary amorphous alloy simultaneously improve the plasticity and glass-forming ability? *Acta Mater.* **61**, 6597-6608 (2013).

31   Alix-Williams, D. D. & Falk, M. L. Shear band broadening in simulated glasses. *Physical Review E* **98**, 053002 (2018).

32   Shimizu, F., Ogata, S. & Li, J. Theory of shear banding in metallic glasses and molecular dynamics calculations. *Mater. Trans.* **48**, 2923-2927 (2007).

33   Șopu, D., Stukowski, A., Stoica, M. & Scudino, S. Atomic-level processes of shear band nucleation in metallic glasses. *Phys. Rev. Lett.* **119**, 195503 (2017).

34   Șopu, D., Scudino, S., Bian, X., Gammer, C. & Eckert, J. Atomic-scale origin of shear band multiplication in heterogeneous metallic glasses. *Scr. Mater.* **178**, 57-61 (2020).

35   Șopu, D. *et al.* From elastic excitations to macroscopic plasticity in metallic glasses. *Applied Materials Today* **22**, 100958 (2021).

36   Șopu, D., Foroughi, A., Stoica, M. & Eckert, J. Brittle-to-Ductile Transition in Metallic Glass Nanowires. *Nano Lett.* **16**, 4467-4471 (2016).

37   Tsamados, M., Tanguy, A., Léonforte, F. & Barrat, J.-L. On the study of local-stress rearrangements during quasi-static plastic shear of a model glass: Do local-stress components contain enough information? *The European Physical Journal E* **26**, 283-293 (2008).

38  Rodney, D., Tanguy, A. & Vandembroucq, D. Modeling the mechanics of amorphous solids at different length scale and time scale. *Modelling and Simulation in Materials Science and Engineering* **19**, 083001 (2011).

39  Cao, P., Short, M. P. & Yip, S. Potential energy landscape activations governing plastic flows in glass rheology. *Proc. Natl. Acad. Sci. USA* **116**, 18790-18797 (2019).

40  Zhang, S., Liu, C., Fan, Y., Yang, Y. & Guan, P. Soft-mode parameter as an indicator for the activation energy spectra in metallic glass. *The Journal of Physical Chemistry Letters* **11**, 2781-2787 (2020).

41  Zhang, S., Wang, W. & Guan, P. Dynamic crossover in metallic glass nanoparticles. *Chin. Phys. Lett.* **38**, 016802 (2021).

42  Fan, Y., Iwashita, T. & Egami, T. Energy landscape-driven non-equilibrium evolution of inherent structure in disordered material. *Nat. Commun.* **8**, 15417 (2017).

43  Fan, Y., Iwashita, T. & Egami, T. Crossover from localized to cascade relaxations in metallic glasses. *Phys. Rev. Lett.* **115**, 045501 (2015).

44  Xu, B., Falk, M. L., Li, J. & Kong, L. Predicting shear transformation events in metallic glasses. *Phys. Rev. Lett.* **120**, 125503 (2018).

45  Xu, B., Falk, M. L., Patinet, S. & Guan, P. Atomic nonaffinity as a predictor of plasticity in amorphous solids. *Physical Review Materials* **5**, 025603 (2021).

46  Liu, C. & Fan, Y. Emergent fractal energy landscape as the origin of stress-accelerated dynamics in amorphous solids. *Phys. Rev. Lett.* **127**, 215502 (2021).

47  Ding, J., Patinet, S., Falk, M. L., Cheng, Y. & Ma, E. Soft spots and their structural signature in a metallic glass. *Proc. Natl. Acad. Sci. USA* **111**, 14052-14056 (2014).

48  Shang, B., Wang, W., Greer, A. L. & Guan, P. Atomistic modelling of thermal-cycling rejuvenation in metallic glasses. *Acta Mater.* **213**, 116952 (2021).

49  Zhu, T., Li, J. & Yip, S. Atomistic study of dislocation loop emission from a crack tip. *Phys. Rev. Lett.* **93**, 025503 (2004).

50  Bitzek, E. & Gumbsch, P. Mechanisms of dislocation multiplication at crack tips. *Acta Mater.* **61**, 1394-1403 (2013).

51  Farkas, D., Duranduru, M., Curtin, W. & Ribbens, C. Multiple-dislocation emission from the crack tip in the ductile fracture of Al. *Philosophical Magazine A* **81**, 1241-1255 (2001).

52  Bei, H., Gao, Y., Shim, S., George, E. P. & Pharr, G. M. Strength differences arising from homogeneous versus heterogeneous dislocation nucleation. *Physical Review B* **77**, 060103 (2008).

53  Cotterell, B. & Rice, J. R. Slightly curved or kinked cracks. *International Journal of Fracture* **16**, 155-169 (1980).

54  He, M.-Y. & Hutchinson, J. W. Kinking of a crack out of an interface. *Journal of Applied Mechanics* **56**, 270-278 (1989).